\documentclass[a4paper]{VisionStyle}
\usepackage{epsfig}

\begin{document}

\title{Faint Source Counts from the Off-source fluctuation 
 Analysis on the Deepest Chandra Fields}

\author{Takamitsu Miyaji \and Richard E. Griffiths} 

\institute{
 Department of Physics, Carnegie Mellon University\\
 5000 Forbes Av., Pittsburgh, PA 15213, USA\\
  e-mail: miyaji@astro.phys.cmu.edu, griffith@astro.phys.cmu.edu	
  }

\maketitle 

\begin{abstract}

  We show the results of the fluctuation analysis applied to
the off-source areas from the two 1 Million second {\it Chandra} 
Deep Fields, including our new results on the Chandra Deep 
Field-South (CDF-S) in the 0.5-2 keV band in addition to those 
on the Chandra Deep Field-North (CDF-N), which have already been reported.  
The distribution of the X-ray counts in cells has been compared 
with the expectation from the ${\rm Log}\; N - {\rm Log}\; S$ model 
to constrain the behavior of the source number density down to 
a factor of several lower than the source-detection limit. We 
show that our results are insensitive to the non-uniformity of 
the non X-ray background (NXB). Our results show that the number counts 
in the soft band (0.5-2 [keV]) continue to grow down to 
$S_{x}\sim 7\times 10^{-18}$ ${\rm [erg\,s^{-1}\,cm^{-2}]}$, 
possibly suggesting the emergence of a new population and they 
agree well with a prediction of star forming galaxies.  


\end{abstract}

\section{Introduction}
\label{tmiyaji-F2_sec:sim}
  
  The number densities of X-ray sources as a function of flux
(the so called the  ${\rm Log}\;N - {\rm Log}\;S$ relation) is 
one of the key constraints for models of the X-ray source 
population. Most of the ``Cosmic X-ray Background'' 
(CXRB) intensity has now been resolved with ``Chandra'' 
(\cite{tmiyaji-F2:mush}; \cite{tmiyaji-F2:rosati}; \cite{tmiyaji-F2:br1Ms}) 
to the extent that major uncertainties in the fraction of the CXRB which 
have been resolved into individual sources, i.e, how much of it remains 
to be explained, lie in the absolute intensity of the CXRB and 
field-to-field fluctuations due to cosmic variance.  In terms of the 
origin of the CXRB, the faintest sources in the {\it Chandra} Deep Fields 
are becoming less interesting. However, constraints on number counts at the  
faintest possible fluxes provide a new view on the nature 
and evolution of the X-ray emitting sources.

 Fluctuation analysis is a strong tool for constraining the number 
densities of sources which are fainter than the source detection 
limit. This technique, which has been pioneered in radio astronomy 
(e.g. \cite{tmiyaji-F2:condon}), has also been successfully applied 
to the X-ray data from previous missions
(e.g. \cite{tmiyaji-F2:hh}; \cite{tmiyaji-F2:has93},
\cite{tmiyaji-F2:geo93}; \cite{tmiyaji-F2:gendreau},
\cite{tmiyaji-F2:yamashita}; \cite{tmiyaji-F2:perri}). 
The source counts inferred from these analyses have turned out to 
be consistent with those from resolved sources in deeper observations
later. Initial results of our work on CDF-N has been reported in 
\cite*{tmiyaji-F2:miy02} (hereafter, MG02).  
In these proceedings, we also report our new 
result on the 0.5-2 keV result on CDF-S, which gave a consistent
result, but more tightly constrained. 

We also discuss a number of technical issues which have not 
been explained in MG02.

\section{Image Simulation and Statistics}
\label{tmiyaji-F2_sec:simstat}

 The analysis has been made by comparing the count distribution of 
$4\arcsec \times 4\arcsec$ image cells ($P(D)$ distribution)
in the real data and the model. Because the point spread 
function (PSF) and exposure varies over the ACIS field of view (FOV), 
analytical modeling of the P(D) distribution cannot be used. Also 
because the probability distribution of the $P(D)$ curve for a 
given model involves the  fluctuations of  both photon counts 
and source counts, it does not follow the simple Poisson statistics. 
Thanks to the rapid improvements in computing 
power in recent years, massive Monte-Carlo simulations can be 
made on our desktops for (1) modeling the P(D) distribution, 
and (2) searching for the 90\% confidence range of the  
${\rm Log}\; N - {\rm Log}\; S$ relation.

 For this purpose, we have developed an image simulator 
({\em qimsim}), which can quickly simulate the summed image 
from multiple pointings of Chandra observations and allows one 
to analyze the statistics of simulated images. {\em Qimsim} 
generates randomly placed point sources for a given 
${\rm Log}\; N - {\rm Log}\; S$ model. For each source generated at a 
sky position and for each pointing, it generates events based on
the exposure map and the off-axis angle from the pointing direction.
This is repeated for all the pointings used for the summed image.
We assume that the background events are dominated by non X-ray
(particle) background (NXB). The NXB is generated using a separate 
``NXB distribution map'' (see below) in such a way that the total 
of the simulated source and the NXB counts is equal to that of 
the real data. 

\begin{figure}[ht]
  \begin{center}
    \epsfig{file=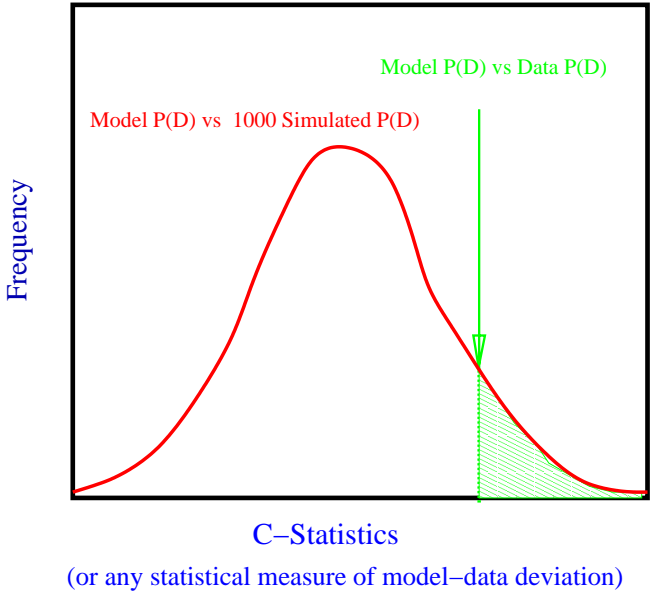,width=\hsize}
  \end{center}
\caption[]{
  The assessment of acceptability of a 
  ${\rm Log}\; N - {\rm Log}\; S$ model using the Monte-Carlo
  simulations is illustrated. The solid curve is the distribution
  of $C_{\rm sim}$ (see text) for the 1000 simulations. The spread
  of $C_{\rm sim}$ is caused by the statistical fluctuations of 
  both the X-ray sources and the detected events. If $C_{\rm sim}$ 
  exceeds $C_{\rm obs}$ in 10\% of the simulations 
  or more (hatched area), we have accepted the model and included
  in drawing the 90\% confidence fluctuation ``horn''.    
   }  	  
\label{tmiyaji-F2_fig:fig1}
\end{figure}

 To assess the significance of the difference 
between the model $P(D)$ (the mean of the $P(D)$ 
curves from the 1000 simulations) and that of the 
observation,  we have used Monte-Carlo simulations. 
For this purpose, we can use any quantity measuring
the difference between the model and the data. For this purpose,
we use the Caster modification of the Cash-C statistics
(\cite{tmiyaji-F2:xspec}; \cite{tmiyaji-F2:cash}; see also
MG02). The acceptance probability has been assessed 
as as follows:
\begin{enumerate}
\item Calculate the distribution of the 1000 $C_{\rm sim}$ values between 
  the model $P(D)$ and the 1000 simulated $P(D)$ distributions derived 
  from the same set of model parameters.
\item Calculate $C_{\rm obs}$ between the data and the model.
\item The fraction of the $C_{\rm sim}$ values which exceed 
  $C_{\rm obs}$ is taken as the model acceptance probability.
\item If this fraction is $>0.1$, we accept it. From the points
  in the model parameter space which have been accepted, we 
  calculate the the 90\% confidence ``horn''. 
\end{enumerate}
This procedure is illustrated in Fig. \ref{tmiyaji-F2_fig:fig1}.

 We plan to use the simulator for further studies involving 
angular correlation functions.

\section{Fluctuation Results, AGNs and Galaxies}
\label{tmiyaji-F2_sec:res}

\subsection{Results}

\begin{figure}[ht]
\begin{center}
\epsfig{file=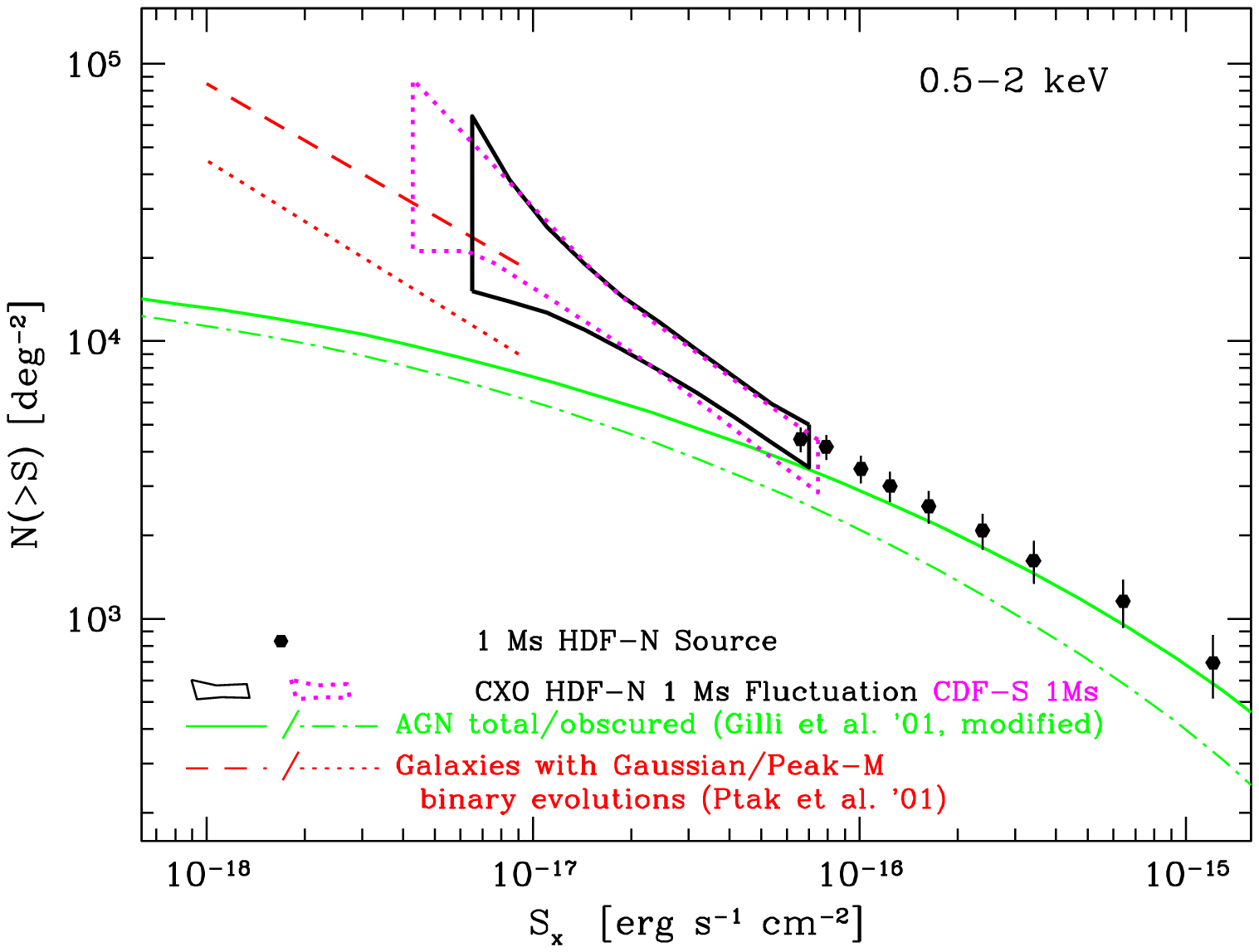,width=\hsize}
\end{center}
\caption[]{The derived ${\rm Log}\;N - {\rm Log}\;S$ relations 
(90\% confidence ``horns'') in the soft band in the CDF North and 
South are compared 
with AGN and galaxy number count models. 

Predictions from the population systhesis model (B) by 
\cite*{tmiyaji-F2:gilli} for the total AGN population and the obscured 
($N_{\rm H}>10^{22}$ $[cm^{-2}]$). Predicted number counts of galaxies 
using the cosmic star-formation history and two models (Gaussian and Peak-M) 
of evolution of X-ray binaries by \cite*{tmiyaji-F2:ptak} are also plotted. 
If mode B of \cite*{tmiyaji-F2:gilli} represents the correct behavior of 
absorbed and unabsorbed AGNs, the fluctuation constraints suggest the 
emergence of an additional population, probably from these galaxies.
}\label{tmiyaji-F2_fig:fig2}
\end{figure}

 As we see in MG02, we could only obtain loose constraints 
on the 2-10 keV band for CDF-N, mainly because of the 
high NXB and limited area of analysis. In these proceedings, we 
concentrate our discussion on the 0.5-2 keV band. 
Fig. \ref{tmiyaji-F2_fig:fig2} shows the 0.5-2 keV 
${\rm Log}\; N - {\rm Log}\; S$ relations for the CDF-N fluctuation, 
CDF-N resolved sources (MG02) and the new fluctuation result from CDF-S.
\footnote{Machine readable ASCII tables of the fluctuation 
horns for our analysis on CDF-N/S can be found at 
http://astrophysics.phys.cmu.edu/$\sim$deepxray/tables.}
The method of the CDF-S analysis is essentially the same as that
of CDF-N reported in MG02. Since the pointing directions 
are much more concentrated in the CDF-S data than the CDF-N case, 
we used the single circle of $5\farcm2$ around the exposure-weighted 
mean pointing direction for our analysis. We have also included the 
{\em resolved source} 
constraint at $S_{\rm x}= 9\;10^{-17}$ ${\rm [erg\,s^{-1}\,cm^{-2}]}$
from the resolved source count 
and the 1$\sigma$ error from \cite*{tmiyaji-F2:rosati} (see 
MG02 for the detailed explanation on incorporating the 
{\em resolved source constraint} into analysis). 
Fig. \ref{tmiyaji-F2_fig:fig2} shows that the faint number 
counts of CDF-N and CDF-S are consistent with each other, but
CDF-S gave a better constraint, probably because we can utilize
a larger area for analysis.  
    
We have overplotted the 0.5-2 keV ${\rm Log}\;N - {\rm Log}\;S$ prediction 
from the AGN population synthesis model composed of unabsorbed and absorbed 
AGNs based on by \cite*{tmiyaji-F2:gilli} (model B). The plotted 
model have been slightly modified from the original as described in 
\cite*{tmiyaji-F2:rosati}. The curves for the total and absorbed 
($N_{\rm H}>22$  ${\rm [cm^{-2}]}$) AGN populations are plotted.  
Fig.\ref{tmiyaji-F2_fig:fig2} shows that the AGN counts drop below 
$S_{\rm x}\sim 10^{-16}$ ${\rm [erg\,s^{-1}\,cm^{-2}]}$ even if the 
emergence of the obscured AGN population 
is taken into account,  while the source counts from the fluctuation 
analysis continue to grow.  If their model B closely represents the 
true behavior of the SXLF of the AGN population, we are probably seeing 
the emergence of a new population of faint X-ray sources.  This excess 
may be contributed by AGNs with low intrinsic luminosities 
(at $L_{\rm x}\la 10^{41.5}$ ${\rm [erg\,s^{-1}]}$) and/or X-rays from 
star-formation activities through supernova remnants and low-/high- mass 
X-ray binaries. In view of this, we have also overplotted 
a model prediction of X-ray number counts based on the cosmic star-formation 
rate \cite*{tmiyaji-F2:ptak} for the two models of binary evolution 
(Gaussian and Peak-M) from \cite*{tmiyaji-F2:ghosh}.  Within the 
uncertainties in the AGN source counts and model construction, these 
predictions gave approximately the number counts from our 
fluctuation analysis.

\subsection{Sensitivity to Non-Uniform NXB}

\begin{figure}[ht]
  \begin{center}
    \epsfig{file=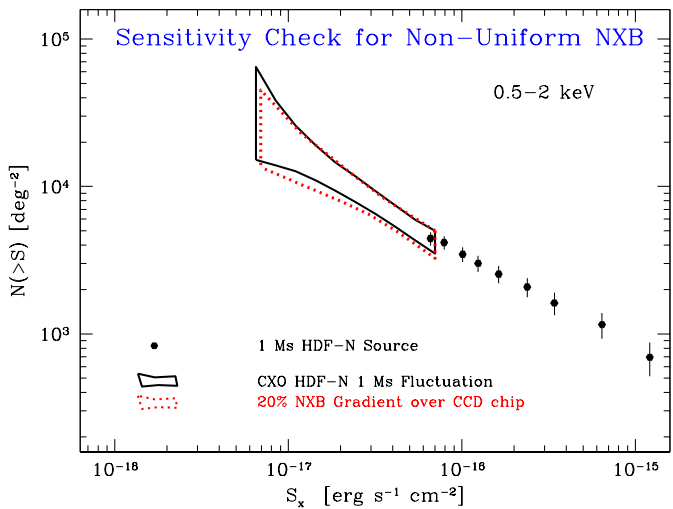,width=\hsize}
  \end{center}
\caption[]{
   Sensitivity to the NXB non uniformity has been assessed. 
  The solid curve assumes that the NXB is uniform over the active
  region of the CCD and dotted curve assumes that the NXB has 
  a 20\% gradient across the ACIS-I CCD chips. The comparison 
  of these two shows that our analysis is
  not sensitive to the non-uniform NXB level.}  	  
\label{tmiyaji-F2_fig:fig3}
\end{figure}

It has been reported that the non X-ray background (NXB) 
varies over the ACIS CCD-chip at the level of $\sim 20-30\%$ across 
the ACIS-I chip for $E>5$ keV, while the variation is much 
less for lower energies.
\footnote{See http://cxc.harvard.edu/contrib/maxim/bg/index.html}.
While we treated NXB uniform over the active region in our analysis, 
one may worry that this NXB non uniformity could  cause 
an overestimate of the derived ${\rm Log}\;N - {\rm Log}\;S$ relation. 
Thus we have also made the analysis assuming that the NXB has a 
20\% gradient along the CHIP-Y coordinate by adjusting the NXB 
distribution map. The results for the uniform and ``slope'' NXB are 
compared in Fig. \ref{tmiyaji-F2_fig:fig3}.  Fig. \ref{tmiyaji-F2_fig:fig3}
shows that the systematic error caused by this effect is much smaller
than statistical errors.

\section{Summary}

\begin{enumerate}
\item Fluctuation analyses have been made on the two 1 Ms Chandra 
   observations on the CDF-North and South in order to constrain the 
   behavior of the ${\rm Log}\;N-{\rm Log}\;S$ at fluxes fainter than 
   the source detection
   limit. We also show our new results on CDF-S for the soft band.
\item We have used Monte-Carlo simulations to obtain model-predicted 
    $P(D)$ curves and evaluate the 90\% confidence constraints.
\item The number counts in the soft band (0.5-2 [keV]) continues
   to grow down to $S_{x}\sim 7\times 10^{-18}$ 
    ${\rm [erg\,s^{-1}\,cm^{-2}]}$, suggesting the emergence of 
    a new population, possibly X-rays from star-formation activity. 
\item The impact of non uniformity of the non X-ray background 
    on our ${\rm Log}\;N-{\rm Log}\;S$ is much smaller than the
    statistical errors. 
\end{enumerate}

\begin{acknowledgements}
This work has been made use of data from the {\it Chandra} Data
Archive. We acknowledge the support from NAG-10875 (TM) and the 
NAS8-38252 (CMU subcontract, REG). We also  thank the members of 
the ACIS team for their help with the analysis as well as Bob Warwick,
Xavier Barcons and G\"unther Hasinger for discussions on the statistics 
and encouragements.  

\end{acknowledgements}

\end{document}